\documentclass[twoside]{LCWS11}
\usepackage[latin1]{inputenc}
\usepackage[dvips]{graphicx}
\usepackage[dvips]{epsfig}
\usepackage{color}
\usepackage{wrapfig,rotating}
\usepackage{amssymb,amsmath,array}
\usepackage{cite}
\usepackage{hyperref}

\graphicspath{{images/}}
\usepackage{subfig}

\begin{document}
\title{Finite Elements Magnetic Analysis of~the~CLIC~MDI~Region} 
\author{Antonio Bartalesi$^1$
\vspace{.3cm}\\
1- CERN - TE Department\\
1211 Geneva 23 - Switzerland
}

\maketitle

\begin{abstract}
Considering the current CLIC SiD detector design and the machine parameter $L^{*}$, the final focus quadrupole QD0 will be placed \emph{inside} the experiment itself. This configuration is very challenging from an integration point of view.\\%
Among several other aspects, the iron-dominated QD0 will need an active magnetic shielding to avoid undesired interactions with the magnetic field generated by the main solenoid of the detector.\\%
This shielding will be provided by a superconducting anti-solenoid, and this paper aims to describe the method used to design such device, the results obtained and the issues still to be solved.
\end{abstract}
\section{Introduction}
\begin{figure}[bh]
\centering
\includegraphics[height=0.4\textheight]{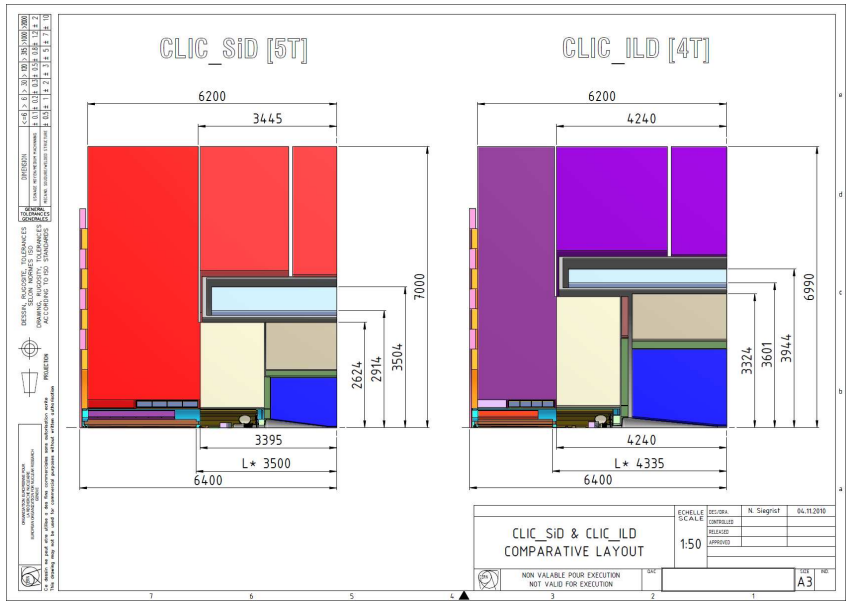} 
\caption{The SiD and ILD designs overview}
\label{fig_sid_ild}
\end{figure}

There are two proposed detectors for CLIC: the SiD and the ILD designs, both represented in figure~\ref{fig_sid_ild}. The SiD~\cite{MSLCWS:2011} is considered to be the most challenging one, due to its higher magnetic field of 5 T at the Interaction Point (IP), therefore the following study focuses on the SiD design only.%
\\%
With an $L^{*}$ value of 3.5 m, the normal conducting final focus lies \emph{inside} the boundaries of the detector as shown in figure~\ref{fig_sid_layout}. In such a compact configuration, QD0 needs to be magnetically shielded, so it was proposed since 2009~\cite{detlef:2009} a superconducting \emph{anti-solenoid} to expel the magnetic field from the region reserved to the final focus.\\%
This anti-solenoid must then be dimensioned and integrated in the compact layout of the SiD detector: in this paper the process followed to design the anti-solenoid, the results obtained and the issues still open are illustrated.
\subsection{The CLiC SiD experiment and the MDI}
\begin{figure}\centering
\includegraphics[scale=1]{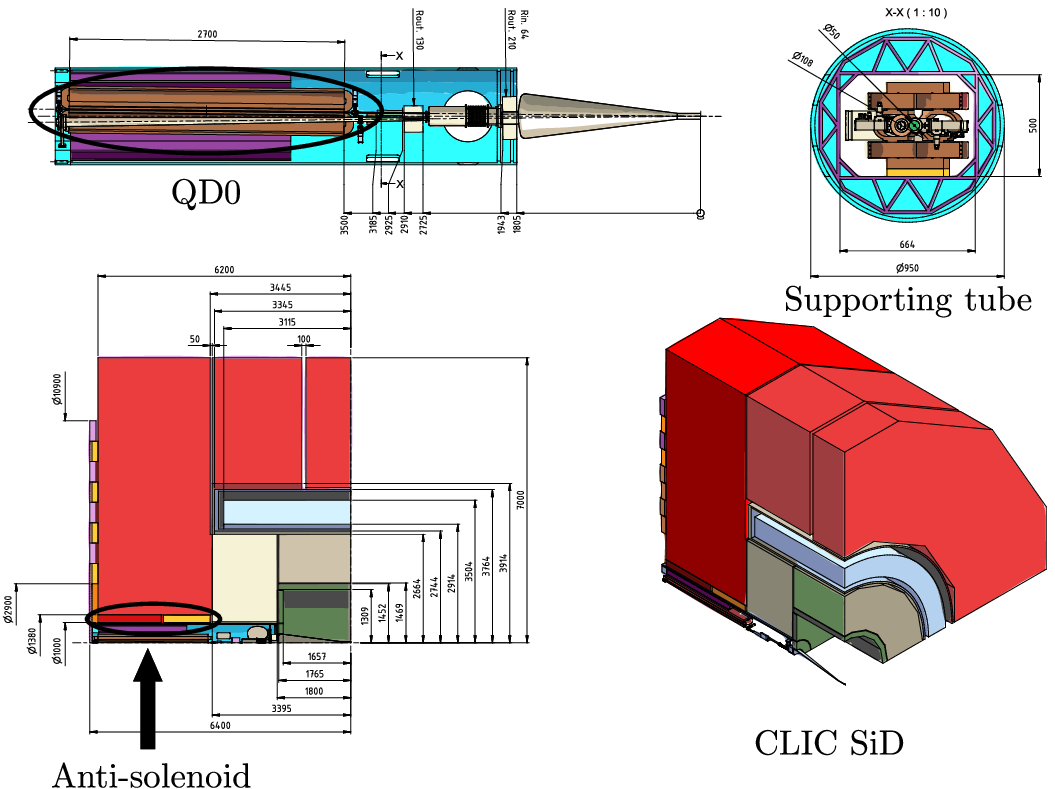} 
\caption{The SiD layout}
\label{fig_sid_layout}
\end{figure}%
In the SiD MDI region many components are required for the most different reasons, and figure~\ref{fig_sid_layout} shows the most relevant to this study. In the magnetic simulations, the following items were \emph{always} modelled: the main solenoid, the flux-return yoke and the anti-solenoid. QD0 was included only in 3D analyses, while any other element was considered as non magnetic and not included in the simulations, just considered for the integration check.
\subsection{Starting point and objectives of the analysis}
\begin{figure}
\begin{center}
\subfloat[Antisolenoid concept in ILD]{\includegraphics[height=0.3\textheight]{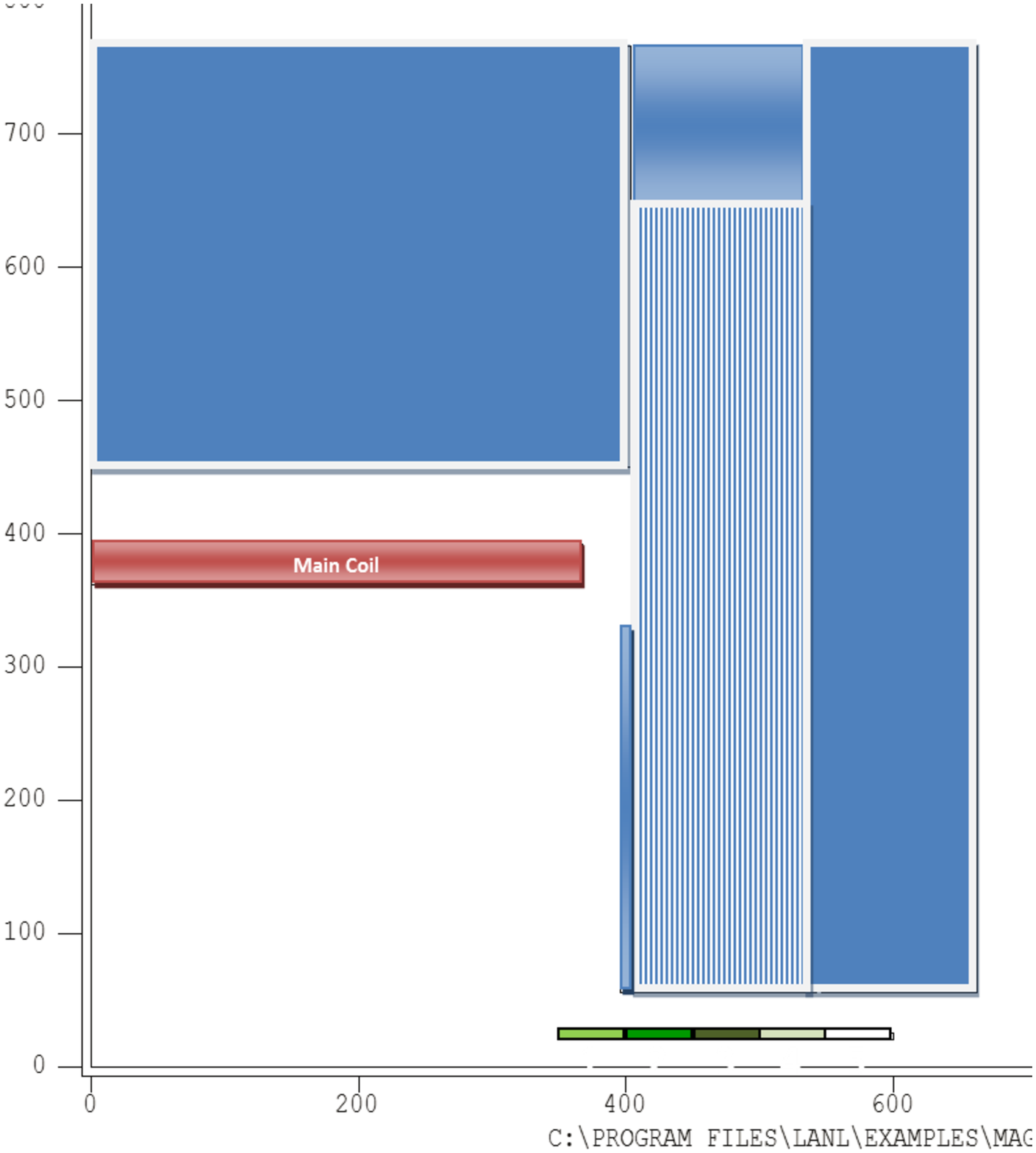}
\label{sfig_swob}}
\subfloat[Flux lines in the SiD with its anti-solenoid]{\includegraphics[height=0.3\textheight]{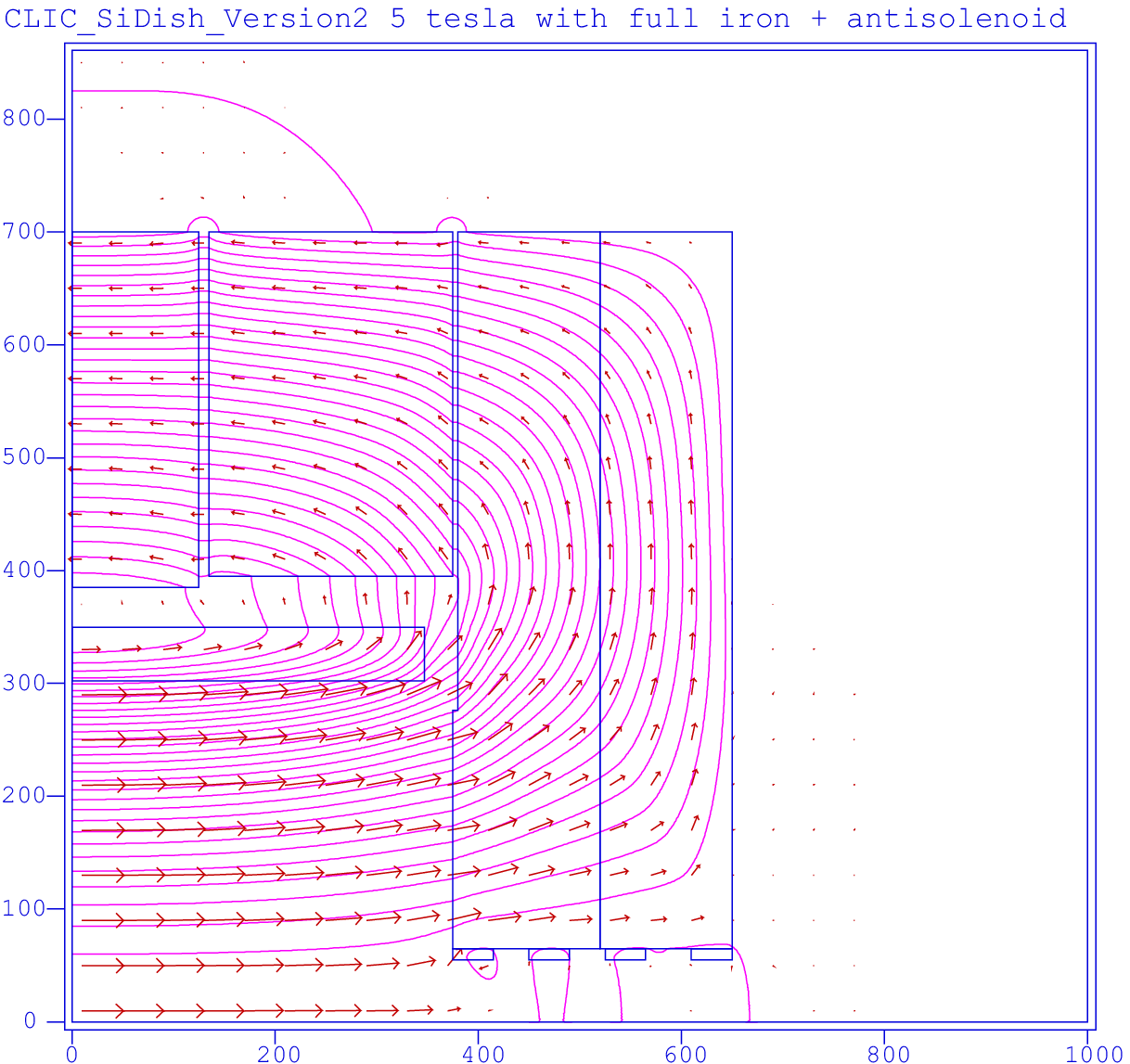}
\label{sfig_hubert}}
\end{center}
\caption{The first anti-solenoid layouts}
\label{fig_swob}
\end{figure}%
Figure~\ref{sfig_swob} shows the first anti-solenoid layout proposed for the ILD detector~\cite{detlef:2009}, while figure~\ref{sfig_hubert} illustrates the flux lines for the first version of the SiD detector with its anti-solenoid. Such configurations were used as concepts for this analysis, since shape and currents of the new anti-solenoid were redefined from scratch.\\%
Summarising, the requirements for the anti-solenoid are:
\begin{enumerate}
\item To provide the correct magnetic shielding to the final focus.
\item To be magnetically compatible with the beam dynamics.
\item To be integrated in the Machine Detector Interface (MDI) region.
\end{enumerate}
The number of coils, their dimensions and currents were set as the output of this study.
\section{The two-dimensional model}
As a first approximation, a 2D axial-symmetric FEM model was considered the best choice to represent the MDI region. In fact, even if QD0 could not be included in this kind of simulation, it was possible to use such model to reduce the magnetic field in the final focus region and evaluate both the forces and the stresses acting on the anti-solenoid. However, since a 3D simulation was considered fundamental, it was planned and performed later.%
\begin{figure}
\begin{center}
\subfloat[First version (ANSYS)]{\includegraphics[scale=1]{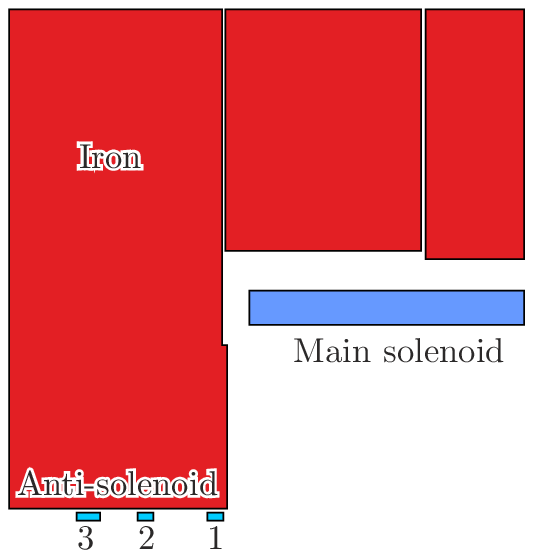}
\label{sfig_ansys_scheme}}
\subfloat[Latest version (Opera)]{\includegraphics[scale=1]{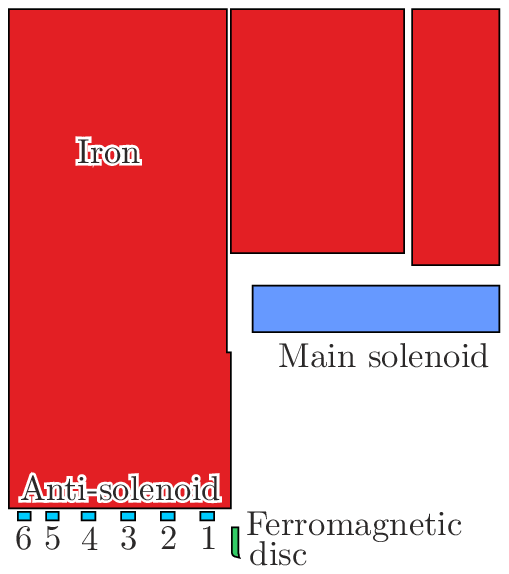}
\label{sfig_opera_scheme}}
\end{center}
\caption{Two finite element models schemes}
\label{fig_ansys}
\end{figure}\\%
The first model was set up using Ansys by Ansol, mainly to evaluate forces and stresses after performing a magnetic simulation. Once found out that magnetic precision was more relevant than stress calculation, that software was abandoned in favour of Opera2D by Vector Fields. Figure~\ref{fig_ansys} shows the very first and the very last designs proposed.\\%
The first anti-solenoid was designed for an $L^{*}$ of 3.8 m, and consisted in 3 superconducting coils, the current densities were 1.14$\cdot 10^8$ A/m\textsuperscript{2}, 3.67$\cdot 10^7$ A/m\textsuperscript{2} and 6.00$\cdot 10^6$ A/m\textsuperscript{2} for coil 1, 2 and 3 respectively. Results in terms of magnetic field along the beam line are shown in figures~\ref{sfig_beamline_ansys} and~\ref{sfig_qd0_ansys}, while the maximum axial force was 7.0$\cdot 10^6$ N, on coil 1.
\subsection{Design evolution of the 2D model}
The first magnetic results obtained were satisfying, but many integration issues still needed a solution. The most important changes that occurred in the evolution of the MDI were the setting of $L^{*}$ to 3.5 m with the consequent detector re-dimensioning; the necessity to reserve a space of at least 40 mm around the anti-solenoid coils for their cryostat, and finally the setting of the maximum current density for the anti-solenoid to 8$\cdot 10^7$ A/m\textsuperscript{2}.\\%
To improve the shielding, anti-solenoids made by different numbers of coils were studied, and also the coils shapes were continuously changed. At some point, it was introduced a ferromagnetic disc to be placed in front of the QD0 region, as shown in figure~\ref{sfig_opera_scheme}, and it was also defined a numeric routine to iteratively find the coil currents which minimize the remaining magnetic field in the QD0 region and on the beam axis.
\subsection{Results of the 2D models and conclusions}%
\begin{figure}[tp]
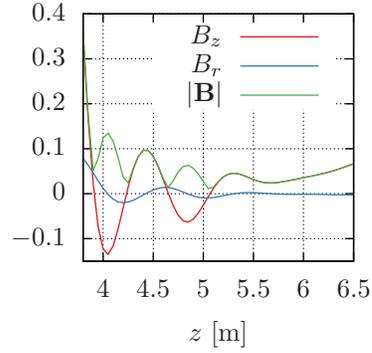
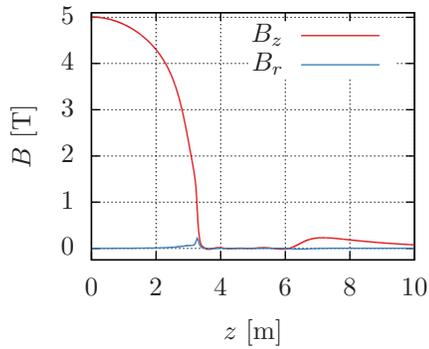
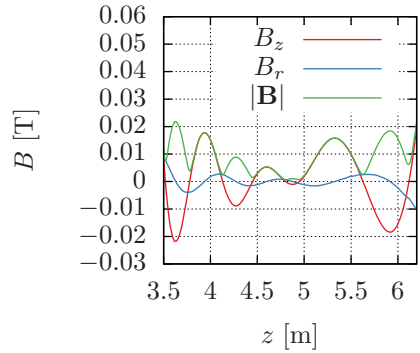
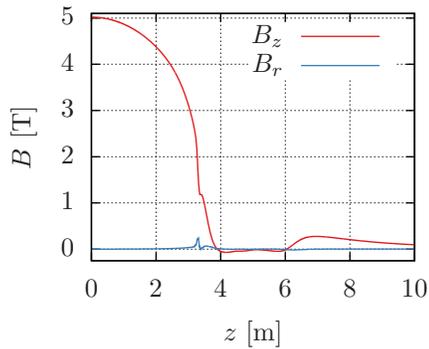
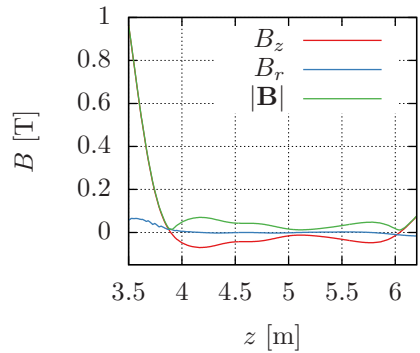

\begin{center}
\subfloat[From the IP to 10 m (Ansys)]{\input{images/Ansys_Bz_Br.tex}
\label{sfig_beamline_ansys}}
\subfloat[In the QD0 region (Ansys)]{\input{images/Ansys_Bz_Br_Bmod_zoom.tex}
\label{sfig_qd0_ansys}}
\end{center}
\begin{center}
\subfloat[From the IP to 10 m (Opera, best result)]{\input{images/Opera_best_Bz_Br.tex}
\label{sfig_beamline_opera_b}}
\subfloat[In the QD0 region (Opera, best result)]{\input{images/Opera_best_Bz_Br_Bmod_zoom.tex}
\label{sfig_qd0_opera_b}}
\end{center}
\begin{center}
\subfloat[From the IP to 10 m (Opera, integrated)]{\input{images/Opera_Bz_Br.tex}
\label{sfig_beamline_opera}}
\subfloat[In the QD0 region (Opera, integrated)]{\input{images/Opera_Bz_Br_Bmod_zoom.tex}
\label{sfig_qd0_opera}}
\caption{Axial, radial and overall magnetic field on the beamline}
\end{center}
\label{fig_Bz_Br_ansys}
\end{figure}%
Figures~\ref{sfig_beamline_opera_b} and~\ref{sfig_qd0_opera_b} show the best results ever obtained in terms of magnetic field with a 2D model. However, due to integration issues, that design was discarded in favour of a less performing one, which gave the results represented in figures~\ref{sfig_beamline_opera} and~\ref{sfig_qd0_opera}.\\%
Once reached this stage, the integration issues were considered solved, the remaining magnetic field minimized and the forces under control. It was then clear that only a 3D simulation could add some more information, and in particular a 3D simulation was the only way to introduce the QD0 magnet in the analysis and study its magnetic interactions with both the detector main solenoid and the anti-solenoid.
\section{The three-dimensional model}
\begin{figure}
\begin{center}
\subfloat[3D model overview]{\includegraphics[width=0.49\textwidth]{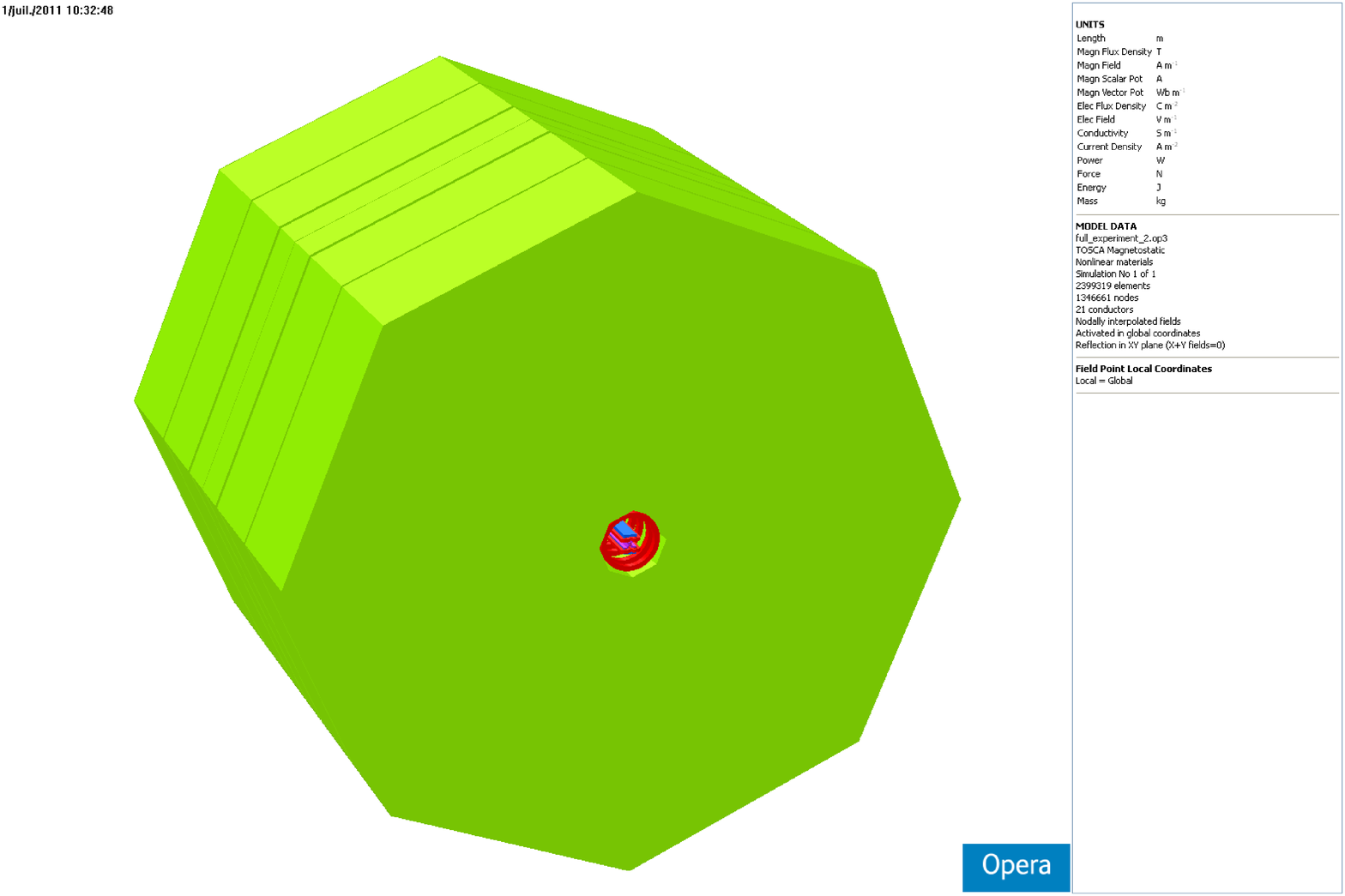}
\label{sfig_3D_over}}
\subfloat[Zoom on the QD0 and its anti-solenoid]{\includegraphics[width=0.49\textwidth]{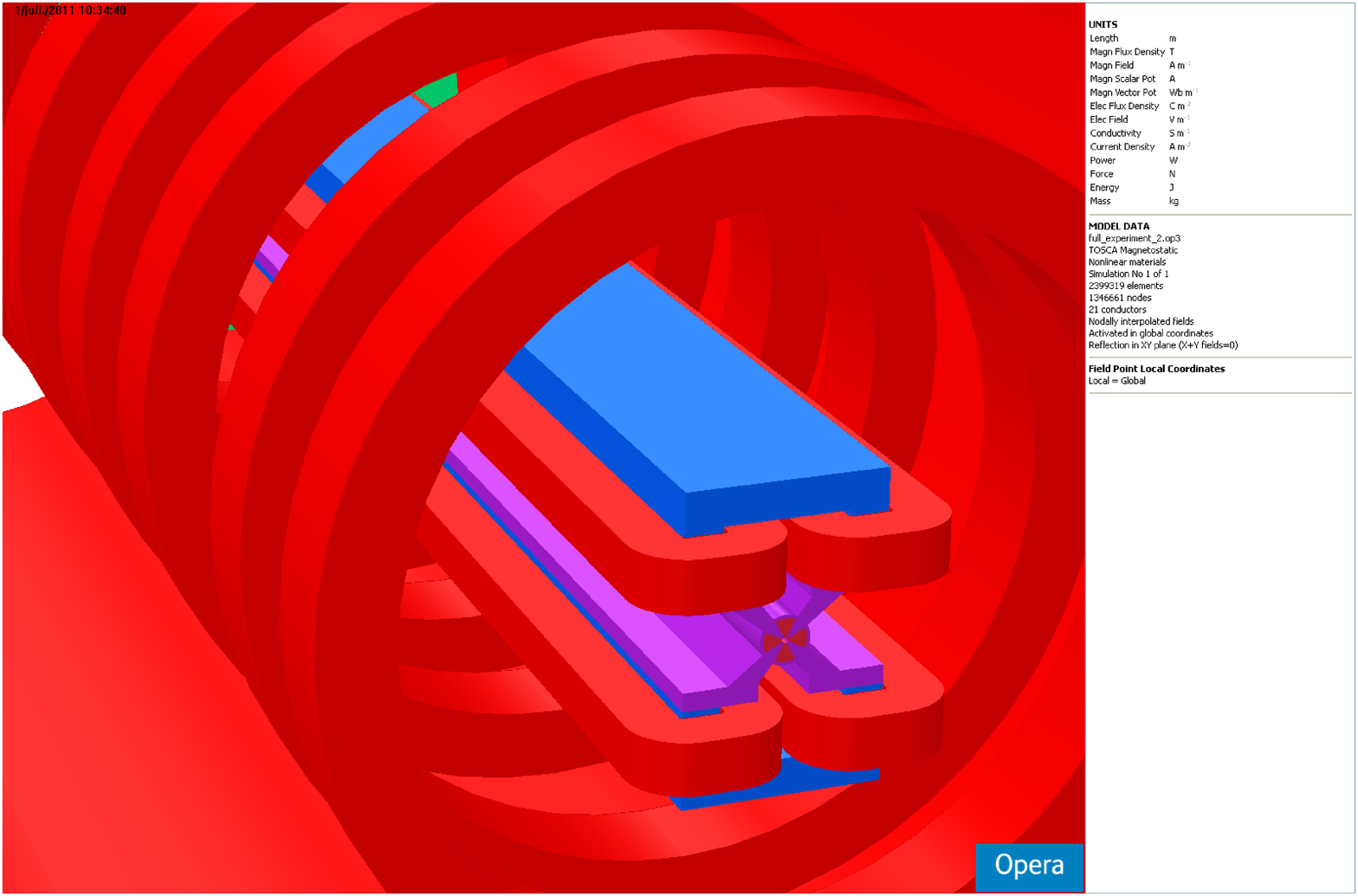}
\label{sfig_3D_QD0}}
\end{center}
\caption{The 3D model}
\label{fig_3D_mod}
\end{figure}

The main challenge of the 3D model is the scale difference of the various components (Figure~\ref{sfig_3D_over}). In the same model there are in fact objects as big as the iron end-cap, with a radius of 7 m, and object as detailed as the final focus, with an aperture radius of just 4.125 mm. Furthermore, it was possible to use only one plane of symmetry to simplify the model, which consisted in a half of the detector plus a full QD0 and an anti-solenoid (Figure~\ref{sfig_3D_QD0}).\\
The magnetic simulation included non linear iron regions, coils and even the permanent magnets blocks which are proposed in the QD0 hybrid design~\cite{QD0LCWS:2011}.
\subsection{Design evolution of the 3D model}
The anti-solenoid defined during the previous 2D analyses didn't perform as before once QD0 was introduced, and too much magnetic field was attracted by its poles and yokes. Then it was necessary to adapt the currents and also reposition slightly some of the anti-solenoid coils, to shield the magnet properly. This was done with a trial and error procedure, since the routine that was set up for the 2D model was based on the superposition of effects and became unusable due to the non linear behaviour of QD0. Finally, the ferromagnetic disc effect was overcome by the presence of the magnet, so it was removed.
\subsection{Results of the 3D model}%
\begin{figure}[t]
\begin{center}
\subfloat[Field on the beamline from the IP to 10 m]{\input{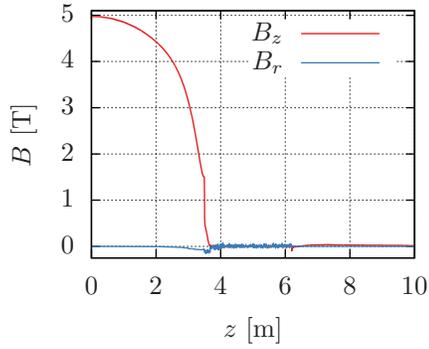}
\label{sfig_3D_BL}}
\subfloat[Field on the beamline in the QD0 region]{\input{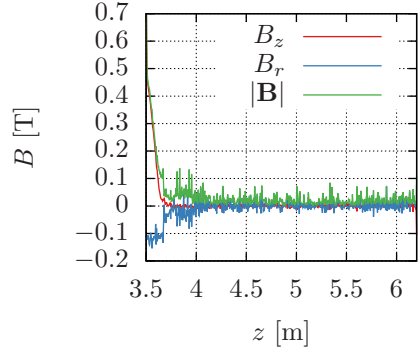}
\label{sfig_3D_BL_zoom}}
\end{center}
\begin{center}
\subfloat[QD0 gradient at 1 mm radius]{\input{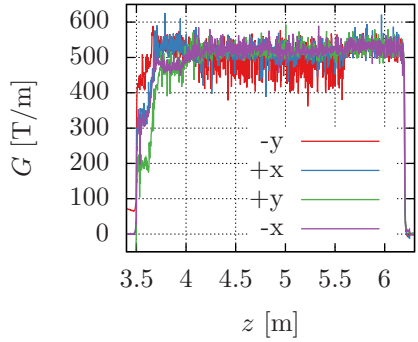}
\label{sfig_grad}}
\subfloat[QD0 gradient at 1 mm radius in the best case]{\input{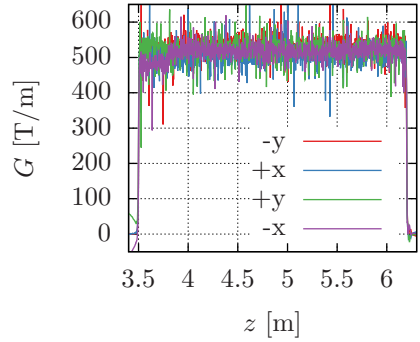}
\label{sfig_grad_good}}
\end{center}
\label{fig_Beee}
\caption{The 3D results}
\end{figure}
Magnetic field on the beam axis was highly reduced as the ferromagnetic QD0 \emph{shielded} the beam line. Figures~\ref{sfig_3D_BL} and~\ref{sfig_3D_BL_zoom} show the remaining field on the beam line.\\
However, since QD0 was simulated, it was also possible to evaluate directly its performance in terms of gradient developed. Figure~\ref{sfig_grad} shows the gradient calculated at 1 mm from the magnetic axis in four different directions, labelled coherently with the beam coordinate system. Some losses are present in the first part (the expected gradient was 535 T/m).\\
Forces on the anti-solenoid were the same as the first 2D analysis and forces on the QD0 were present and highly related to the shielding efficiency. With the best integrated design the maximum force on the QD0 was roughly 5 KN, in the axial direction towards the IP.
\subsection{A last non integrated version}
Since the QD0 performance was still not satisfying, it was decided to put aside integration and try to define a better anti-solenoid. The latest design investigated was not compatible with the pre-alignment system~\cite{PREALLCWS:2011}, since the first coil was closer to the IP, but it was not interfering with the Hadronic Calorimeter or any other system. Figure~\ref{sfig_grad_good} shows the improvements in terms of gradient. Losses were reduced, as the first part (about 30 cm) of QD0 worked at 95\%. The only solution found so far to achieve 100\% of the QD0 gradient was to increase $L^{*}$ by 30 cm.
\subsection{Conclusions and next steps}
This study points out that the anti-solenoid \emph{can} shield the QD0, even if this will require some additional space, or a slightly bigger $L^{*}$. This will also reduce the forces acting on the magnet, while in the anti-solenoid forces are still relevant and instability issues may occur.\\
As next steps, transient analyses, further integration and better FEM models are foreseen.
\bibliographystyle{lcd}
\bibliography{lcd}
\end{document}